\documentclass[aps,pra,showpacs,amsmath,amssymb,twocolumn]{revtex4}

\usepackage{graphicx}% Include figure files
\usepackage{dcolumn}% Align table columns on decimal point
\usepackage{bbm}% bold math

\newcommand{ \mo }[1]{ \hat{ #1 } }

\newcommand{ \sD }{ \mathcal{D} }
\newcommand{ \sH }{ \mathcal{H} }

\newcommand{ \sE }{ \mathbb{E} }

\newcommand{ \Tr }{ \mbox{Tr} }

\newcommand{\be}{\begin{equation}}
\newcommand{\ee}{\end{equation}}
\newcommand{\ba}{\begin{array}}
\newcommand{\ea}{\end{array}}
\newcommand{\bea}{\begin{eqnarray}}
\newcommand{\eea}{\end{eqnarray}}
\newcommand{\n}{\nonumber}

\begin{document}

\title{Scalable quantum field simulations of conditioned systems}

\author{M. R. Hush}
\author{A. R. R. Carvalho}
\affiliation{Department of Quantum Science, Research School of Physics and Engineering, The Australian National University, ACT 0200, Australia}
\author{J. J. Hope}
\affiliation{Australian Centre for Quantum-Atom Optics, Department of Quantum Science, Research School of Physics and Engineering, The Australian National University, ACT 0200, Australia}

\date{\today}

\begin{abstract}
We demonstrate a technique for performing stochastic simulations of conditional master equations. The method is scalable for many quantum field problems and therefore allows first-principles simulations of multimode bosonic fields undergoing continuous measurement, such as those controlled by measurement-based feedback.  As examples, we demonstrate a 53-fold speed increase for the simulation of the feedback cooling of a single trapped particle, and the feedback cooling of a quantum field with 32 modes, which would be impractical using previous brute force methods.
\end{abstract}


\maketitle

\section{Introduction}
The precise generation and control of quantum systems is necessary for any proposed experiment in quantum information and quantum computing, and many potential applications in precision measurement \cite{nielsen00,milburn_pop_book}. It is also necessary for sensitive tests of quantum mechanics and emergent phenomena in quantum physics. Just as it is for classical devices, measurement-based feedback control~\cite{belavkin87,wiseman_milburn_feedback93,wiseman94,doherty99,vanhandel05} is a vital tool for improving the control and stability of quantum systems \cite{morrow02,armen02,geremia03,reiner04,bushev06,felinto06}.  Due to the fact that the size of a Hilbert space grows exponentially with the degrees of freedom of a quantum system, simulating the behaviour of large quantum systems is a difficult process. This makes it hard to model and design feedback for non-trivial quantum systems.

For high-dimensional {\it unconditional} quantum evolution, the most effective ways for direct simulation have been phase space methods using stochastic techniques~\cite{gardiner:01,gardiner:02}. These approaches map the master equation describing the system to a Fokker-Planck equation (FPE) for a quasi-probability distribution. The evolution of this distribution is obtained by considering the average behaviour of a set of stochastic variables, akin to the solution of the Langevin equations corresponding to a FPE. Not all master equations can be simulated efficiently in this fashion using current techniques, but stochastic methods have been used extensively to model low-dimensional quantum optical systems \cite{gardiner:02}, low-dimensional atom optical systems \cite{wiseman:01}, optical~\cite{Drummond:87}, atomic~\cite{PhysRevA.64.053608} and even fermionic quantum fields \cite{Corney:06}.  For example, a single optical mode in an optical cavity can be simulated using the Wigner representation which has 2 degrees of freedom, the amplitude and phase quadrature.  When converted to a set of stochastic differential equations, only 2 real-valued equations are required.  Simulation of $M$ optical modes requires 2$M$ stochastic equations \cite{gardiner:01,gardiner:02}.  This linear scaling with number of modes is contrasted with the exponentially increasing size of the Hilbert space.  In the infinite dimensional limit, quantum fields with $D$ spatial dimensions can be simulated with a $D$-dimensional stochastic partial differential equation, which has been used for first principles calculations in a variety of systems \cite{Drummond:87,PhysRevA.64.053608,Corney:06}. 

Unfortunately, for models of {\it conditional} systems, such as those under continuous monitoring or controlled by measurement-based feedback, an equivalent unravelling of the FPE into low-dimensional stochastic equations cannot be obtained using current techniques. In this paper we develop a method of performing this unravelling, therefore extending the scalability properties of phase space stochastic methods to a whole new class of problems where conditioning is required.

Models of systems undergoing measurement-based feedback require the development of \textit{conditional} master equations with stochastic elements describing the outcome of measurement results~\cite{belavkin87,wiseman_milburn_feedback93,wiseman94,doherty99,bouten07}. This is because the state of the system \textit{correlated to a given measurement record} is required to model the effect of applying feedback control based on that measurement record. Note that the stochasticity introduced here is of a different nature of that obtained from the unravelling of a FPE. While the latter is a fictitious noise used to map the evolution of a distribution in terms of random trajectories, the former is a real noise generated by the measurement process. The dynamics of a conditional master equation can be mapped to the evolution of any corresponding quasi-probability distribution using standard methods, but the resulting equation of motion, which we will call a stochastic Fokker-Planck equation (SFPE), remains stochastic. 

Conditional quantum systems have been simulated using trajectory methods, which reduce the size of the problem by treating the evolution of the conditional density matrix as an average of an ensemble of state vectors undergoing a stochastic process~\cite{carmichael_book1,gambetta05}.  This reduces the dimensionality of the problem from $N^2$ to $N$, where $N$ is the size of the Hilbert space of the quantum system. Unfortunately, $N$ scales as the exponential of the number of degrees of freedom in the system (e.g. the number of qubits, or the number of single particle states of a quantum field), so these methods will never be tractable for truly high dimensional systems.  Thus, some equivalent of the stochastic unravelling of quasi-probability representations must be found that can be applied to conditional quantum systems.  An unravelling has been found for an equation of motion for a classical conditional probability distribution, called the Kushner-Stratonovich equation (KSE) \cite{KSE,gambetta05}.
The resulting low-dimensional stochastic equations had both kinds of noise discussed above: the `fictitious noise' that was introduced so that it would average out to reproduce the diffusion terms of the KSE, and the noise from the KSE itself, which is a function of the actual measurement process. These equations used weighted trajectories, which have also been used in quantum simulations of master equations where the freedom introduced can produce stochastic equations without singularities or instabilities \cite{deuar02}. Unfortunately, mapping a generic conditional master equation to the evolution of a quasi-probability distribution produces a Fokker-Planck equation (FPE) with additional stochastic terms, rather than a KSE.  

In section \ref{sec:technique} we describe the general form of the stochastic technique that simulates the stochastic FPE.  In section  \ref{sec:singleparticle} we apply this method to a low-dimensional example where the FPE can be solved directly, a single trapped particle being cooled by feedback control to the trapping potential.  We use the method on a trapped quantum field in section \ref{sec:quantumfield} to show that a high-dimensional example is still tractable.

\section{Unravelling technique} \label{sec:technique}
We will now demonstrate that a low-dimensional stochastic unravelling of these stochastic FPEs can be achieved at the cost of both introducing weights and simultaneous integration of all members of the ensemble.
Consider a general diffusive conditional master equation~\cite{gisin84,diosiA,diosiB,barc_belav91,wiseman_milburn_homodyne93} 
\begin{equation}
d\rho_c = -\frac{i}{\hbar} [H,\rho_c] + \sum_j \sD[L_j] \rho_c + \sum_j \sH[L_j]dW_j,
\label{me}
\end{equation}
representing the dynamics under the Hamiltonian $H$ and the continuous monitoring of the operators $L_i$. $\sD[L] \rho \equiv L \rho L^\dagger - 1/2(L^\dagger L \rho + \rho L^\dagger L)$ and $\sH[L]\rho \equiv L \rho+\rho L - 2\rho\langle L \rangle$ correspond to the decoherence and to the innovation terms introduced by the measurement, respectively. Stochastic equations will be written in either Stratonovich or Ito forms and will be indicated by the Wiener noises with ($dW^{(s)}$) or without ($dW$) superscript, respectively.

Using a phase space representation \cite{gardiner:02}, this master equation can be converted to a stochastic partial differential equation that is often of the form:
\begin{widetext}
\begin{equation}
d p(\mathbf{x},\mathbf{W}(t),t) = \left(\left\{ - \partial_i A_i + \frac{1}{2}\partial_i[ C_{ik} \partial_{i'}[C_{i'k}]]  + \alpha - \langle{\alpha}\rangle\right\} dt + \left\{ - \partial_i B_{ij} + \beta_j - \langle{\beta}\rangle_j\right\} dW_j^{(s)}\right) p(\mathbf{x},\mathbf{W}(t),t), \label{eq:prob}
\end{equation}
\end{widetext}
where we use Einstein summation notation and suppress functional dependences for brevity. In this and the following equations, the indexes $i$ and $i'$ span the variables in the phase space representation, the index $j$ spans the Linblad operators in Eq.~(\ref{me}), and the index $k$ spans the size of the matrix $C$.  $p$ is the chosen quasi-probability distribution, $\langle{f}\rangle = \int d\mathbf{x} \;p(\mathbf{x}) f(\mathbf{x})$, $\partial_i \equiv \partial/\partial x_i$, and $\mathbf{x}$ and $\mathbf{W}$ are, respectively, the sets of variables describing the system and the Wiener noises associated with the measurement. $A_i$, $B_{ij}$, $C_{ij}$, $\alpha$, $\beta_i$ are functions of $\mathbf{x}$ that are determined by Eq.~(\ref{me}), and the choice of quasi-probability distribution.  For simulations involving measurement-based feedback, these functions may also depend on the distribution $p$, making the equation nonlinear. The first two terms form a Fokker-Planck equation for which standard unravelling techniques are applicable, and the rest arise due to the conditional dynamics.  

We will now show that the following set of weighted stochastic differential equations (WSDE) for the stochastic variables ${x_i}$ and weight $\omega$,
\begin{eqnarray}
dx_i(t) & = & A_i dt + \sum_j B_{ij}(x(t),t) dW_j^{(s)}(t) \nonumber \\
&  & + \sum_k C_{ik}(x(t),t) dV_k^{(s)}(t), \nonumber \\
\frac{d\omega(t)}{\omega} & = & \alpha(x(t),t) dt + \sum_j \beta_j(x(t),t) dW_j^{(s)}(t), \label{eq:WSDE} 
\end{eqnarray}
is a valid unravelling of Eq.~(\ref{eq:prob}). Here, $dW_j$ are real noises corresponding to different actual runs of an experiment and $dV_k$ is a set of artificial noises introduced by the unraveling.  The number of these artificial noises is determined by the shape of the matrix $C$, which does not have to be square and is not uniquely determined.  This is not a unique factorisation of the equation of motion for the quasi-probability distribution $p$.  This can lead to optimisation choices, often called `diffusion gauges', but once that factorisation is chosen, as in Eq.~(\ref{eq:prob}), we find that we must introduce an equivalent number of noises. These increments obey the traditional Ito rules
\begin{equation}
 dW_j dW_{j'}  = \delta_{j j'} dt; \;\; dV_k dV_{k'} = \delta_{k k'} dt; \;\; dV_kdW_j = 0, \label{eq:itorules}
 \end{equation} 
 and we denote the averaging over fictitious noises as $\sE[\circ]$. 
  
Each path is assigned a `weight' $\omega$,  so that observables are calculated using $\sE[\omega f(\textbf{x})]/\sE[\omega]$, where we divide by $\sE[\omega]$ for normalization.  We will use the notation 
\begin{equation}
\overline{f(\textbf{x})} \equiv \sE[\omega f(\textbf{x})]/\sE[\omega] \label{eq:cobs}
\end{equation}
to indicate these weighted averages.  

Using Eq.~(\ref{eq:WSDE}) and the Ito rules (\ref{eq:itorules}), we find that the differentiation rule for the  averages in Eq.~(\ref{eq:cobs}) is given by 
\begin{eqnarray}
d\overline{f(\textbf{x})} & = & \left\{ \overline{\sum_i A_i \partial_i f(\textbf{x})} + \frac{1}{2} \sum_{i i' k} C_{i' k} \partial_{i'} C_{ik} \partial_{i} f(\textbf{x}) \right. \n \\
& & \left. \overline{\alpha f(\textbf{x})} - \overline{\alpha}\,\overline{f(\textbf{x})} \right\} dt \nonumber \\
& & +\sum_j \left\{ \overline{\sum_i B_{ij} \partial_i f(\textbf{x}) } + \overline{\beta_j f(\textbf{x})} - \overline{\beta_j}\, \overline{f(\textbf{x})} \right\}  dW_j^{(s)}.\n \\
\end{eqnarray}
We are now in position to show that the stochastic average $\overline{f(x)}$ coincides with the average $\langle f(x) \rangle$ extracted from the probability distribution. Substituting Eq.~(\ref{eq:prob})~in $d\langle{f}\rangle = \int d\mathbf{x} \, d p(\mathbf{x}) f(\mathbf{x})$, integrating by parts and assuming boundary terms vanish, we get $d\langle{f}\rangle = d \overline{f(x)}$. We have thus shown that moments of a quasiprobability distribution with evolution given by Eq.~(\ref{eq:prob}) are given by the weighted averages of our SDEs (\ref{eq:WSDE}). This means that a class of {\it conditional} master equations for a quantum system with an $N$-dimensional Hilbert space can be simulated by a set of SDEs of size $log(N)$, and we have the central result of this paper. 

\section{Example: Single trapped particle}\label{sec:singleparticle}
As a first example of this new technique we will examine the model for cooling a single particle undergoing a position measurement based feedback derived in~\cite{doherty99}, and extended for non-Gaussian states in~\cite{wilson07}. The conditional master equation for such a system is given by
\begin{equation}
d\rho_c = -i [\hat{H},\rho_c] dt + \gamma \sD[\hat{x}] \rho_c dt + \sqrt{\gamma} \sH[\hat{x}] \rho_c dW, \label{eq:condsingle}
\end{equation}
where $\hat{H} = \hat{x}^2/2 + \hat{p}^2/2 - u(t)\hat x$, and  $u(t) = k_p \Tr[\hat{p} \rho_c]$ is the control signal.
The equivalent SFPE for the Wigner ($\mathcal{W}$) distribution is
\begin{eqnarray}
d\mathcal{W}(x,p,t) & = & \left(\partial_{p} (x-u) - \partial_x p + \frac{\gamma}{2} \partial_{p}^2 \right. \n \\
& & \left. - \gamma((x - \overline{x})^2 - \overline{(x - \overline{x})^2}) \right) \mathcal{W} dt \n \\
& & +  2\sqrt{\gamma} (x - \overline{x}) \mathcal{W} dW^{(s)}(t). \label{eq:Itowigxp}
\end{eqnarray}

We can convert Eq.~(\ref{eq:Itowigxp}) into a set of SDEs using Eq.~(\ref{eq:WSDE}):
\begin{eqnarray}
dx(t) & = & p \;dt, \nonumber \\
dp(t) & = & -(x-u) dt + \sqrt{\gamma} dV^{(s)}, \nonumber \\
\frac{d\omega(t)}{\omega} & = & -2\gamma(x-\overline{x})^2dt + 2\sqrt{\gamma}x dW^{(s)}. \label{eq:wpsdesing}
\end{eqnarray}
The first two equations are the SDEs governing a harmonic oscillator driven by a measurement-induced white noise force. Note that the equation for the weights contains all the information from the innovations term. 

\begin{figure}[htb]
\includegraphics[scale=0.45]{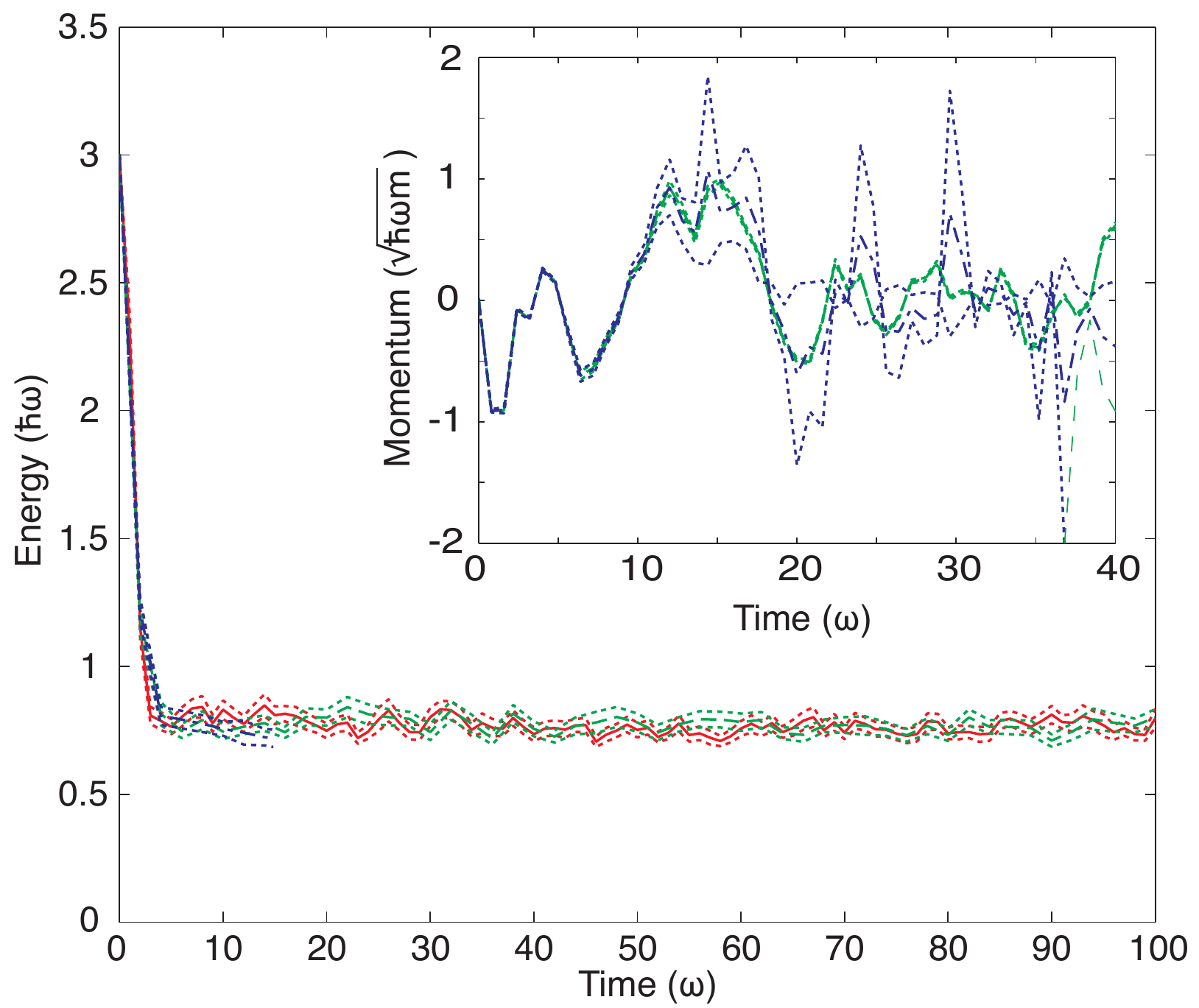}
\caption{(Color online) Energy vs. time for a single particle undergoing measurement-based feedback averaged over 10000 `fictitious' noises and 100 `real' noises. An initial coherent state displaced in position with an initial energy of $3\hbar\omega$ is effectively cooled. We compare simulations using the Wigner phase space method (solid,red), and our WSDEs with (dashed,green) and without (dot-dashed,blue) breeding. Dotted lines correspond to the standard errors. The estimation of the momentum variable becomes rapidly inaccurate without breeding (see inset for momentum evolution over a single `real' noise path). This inaccuracy is fed back into the equations of motion resulting in failure of the integration method. Breeding corrects this sampling problem ensuring convergence to the exact solution for longer times.}
\label{fig:1}
\end{figure}

We can analyse the convergence of this new technique by comparing the solutions of Eqs.~(\ref{eq:Itowigxp}) and (\ref{eq:wpsdesing}) as shown in Fig.\ref{fig:1}.  These simulations were performed with an initial state corresponding to a position-displaced ground state, and $k_p=-1.35$. The simulation was performed using a Mersenne twister based random noise generator to ensure the fictional and real noises remain uncorrelated. The stochastic method converges to the same solution as the Wigner representation over a limited interval due to sampling errors. However, the long-term convergence of these simulations can be enhanced dramatically by using a `breeding' or `branching' technique~\cite{trivedi90}. Trajectories that evolve to give negligible contribution can be ignored in favour of resampling the remaining ones. If a weight is found to be smaller than a chosen tolerance $\epsilon$, i.e. $\omega'_{small}/\langle\omega\rangle < \epsilon$, the memory used to store this path is freed and the path with the largest weight $\omega_{max}$ is resampled. This means the variables of the $\omega_{max}$ path are copied into $\omega'_{small}$, and the $\omega_{max}$ weights are halved such that the calculated observables are still equal within the tolerance of the integration. This increases the effective sampling and the stochastic method converges over the full interval.  Like most numerical techniques if the error tolerance $\epsilon$ is too large, the resampled distribution does not retain all the properties of the original distribution.  To ensure that the breeding technique is convergent, the simulation must be tested by repeated simulations with lower tolerances.  When a lower tolerance is required, a reduced $\epsilon$ must be accompanied by an increased sample size.

The primary advantage of stochastic techniques is that memory requirements scale well for large Hilbert spaces.  For conditional simulations, the dependence of the evolution on expectation values requires simultaneous integration of all paths, so the actual integration is affected by sampling error.  This is in contrast with simulation of traditional master equations, where the sampling error is a purely statistical error in the final averages.  Although simultaneous integration of multiple paths is an increase in memory demand, this is more than compensated by the log(N) memory requirements of the individual paths, indicating that these techniques are still feasible for quantum fields. We can demonstrate this advantageous scaling by considering the multi-particle extension of the single particle problem described in~Eq.(\ref{eq:condsingle}), where we model the evolution of a trapped bosonic quantum field under feedback control.

\section{Example: Trapped quantum field} \label{sec:quantumfield}
The simplest extension of the previous example to a high-dimensional system is to consider the case of a harmonically trapped quantum field where we can control the position of the center of the trap.  For an ideal measurement of the centre of mass motion of the trapped field, we have the following conditional master equation:

\begin{equation}
d\rho_c = -i [\hat{H},\rho_c] dt + \gamma \sD[\hat{X}] \rho_c dt + \sqrt{\gamma} \sH[\hat{X}] \rho_c dW,   \label{eq:condmulti} \\
\end{equation}
where the Hamiltonian is $\hat{H} = \int dx \,  \mo{\psi}^\dag(x) (x^2/2 - \partial_x^2/2 - u(t) x )\mo{\psi}(x) $, $\hat{\psi}(x)$ is the field annihilation operator, and the observable for the centre of mass position of the trapped field is $\hat{X} = \int dx \, x \mo{\psi}^\dag(x) \mo{\psi}(x)$.

We can first convert this equation into a functional positive-P representation~\cite{steel98}, $\mathcal{P}(\phi(x),\xi(x),W(t),t)$, then use the techniques outlined above to convert them to a set of WSDEs:
\begin{eqnarray}          
d\phi(x,t) &=& -iH(x) \phi dt - 2\gamma x (X - \overline{X}) \phi dt \n \\
& & + \sqrt{\gamma} x (i\phi dV^{(s)}_1 + i\phi dV^{(s)}_2 + \phi dW^{(s)} ), \n \\
d\xi(x,t) &=&  -iH(x) \xi dt - 2\gamma x (X - \overline{X}) \xi dt \n \\
& & + \sqrt{\gamma} x (-i\xi dV_1^{(s)} + i\xi dV_2^{(s)} + \xi dW^{(s)} ),\n \\
\frac{d\omega(t)}{\omega} &=& -\gamma(X^{(2)} + (X - \overline{X} )^2) dt + \sqrt{\gamma}  X  dW^{(s)}\n \\
\label{eq:wsdesmulti}
\end{eqnarray}
with
$X  = \int dx \; x \phi(x) \xi(x)$, and 
$X^{(2)}  = \int dx \; x^2 \phi(x) \xi(x)$.

\begin{figure}[htb]
\includegraphics[scale=0.45]{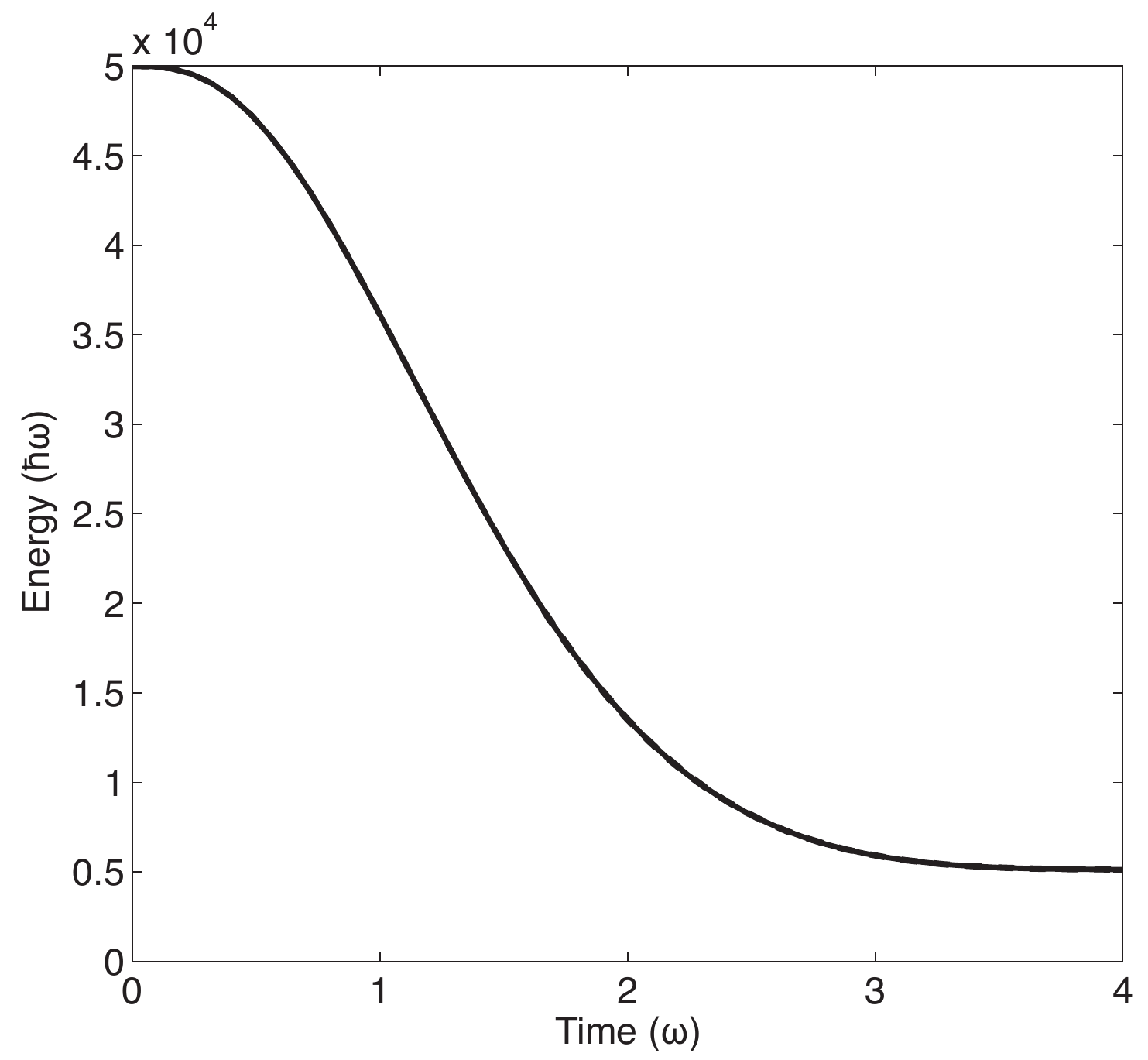}
\caption{Energy for a multimode quantum field calculation using 20 realisations of the WSDEs method with breeding.}\label{fig:2}
\end{figure}

Equations~(\ref{eq:wsdesmulti}) were solved numerically in one dimension with 32 modes and 1000 realisations of the `fictitious noise'.  The same parameters as the single particle calculation were used, and similar cooling behaviour is observed.  The average results of 20 realisations are shown in Fig.~\ref{fig:2}. Each simulation took 6 minutes on a personal computer using the XMDS numerical package \cite{xmds:01}, showing that sizable conditional quantum problems can be computed in reasonable time with this method.

\section{Conclusions}
The stability of all stochastic methods depends strongly on the dynamics of the system, as the simulation is always more efficient when an appropriate basis is used for the quasi-probability representation.  The introduction of a measurement tends to project the system towards eigenstates of that measurement, so the choice of measurement in the system has a strong effect on the stability of any stochastic method based on a given quasi-probability distribution. The stochastic technique presented here will be most efficient when the underlying basis of the representation is a reasonable match for the likely states of the conditioned system.

This paper has described a stochastic method that can simulate conditional quantum systems undergoing feedback. The cooling of a single trapped atom is simulated as an example, and compared to the evolution using a direct simulation of the Wigner function. The stochastic method presented in this paper is 53 times faster to compute, but its real advantages over `brute force' calculations come from its logarithmic scaling with the size of the Hilbert space.  This scaling is demonstrated by the first-principles simulation of a trapped single-dimensional bosonic field undergoing position measurement and feedback, which is a simulation that can only be performed by this new stochastic method.  This technique opens the possibility for exploration of non-trivial quantum systems undergoing feedback control.  

\section*{Acknowledgements}
The authors would like to acknowledge beneficial conversations with Prof.~Howard Wiseman.  This work was Þnancially supported by the Australian 
Research Council Centre of Excellence program.  Numerical simulations were done at the National Computational Infrastructure National Facility.\bibliography{references}

\end{document}